\documentclass[aps, prl, twocolumn, superscriptaddress,numerical,showpacs,longbibliography,floatfix,footnoteinbib]{revtex4-1}
\usepackage{amsmath,amssymb}
\usepackage{graphicx}
\usepackage{bm}
\usepackage{slashed}
\usepackage[usenames,dvipsnames,svgnames,table]{xcolor}
\usepackage[colorlinks=True,linkcolor=blue,citecolor=blue,urlcolor=blue]{hyperref}

\usepackage{soul}

\let\oldciteauthor=\citeauthor
\def\citeauthor#1{\hypersetup{citecolor=black}\oldciteauthor{#1}}

\let\oldciten=\onlinecite
\def\onlinecite#1{\hypersetup{citecolor=blue}\oldciten{#1}}

\let\oldcite=\cite
\def\cite#1{\hypersetup{citecolor=blue}\oldcite{#1}}

\newcommand*\diff{\mathop{}\!\mathrm{d}}
\newcommand{\bra}{\langle}
\newcommand{\ket}{\rangle}

\begin{document}

\title{Axial anomaly generation by domain wall motion in Weyl semimetals}

\author{Julia D.\ Hannukainen}
\affiliation{Department of Physics, KTH Royal Institute of Technology, Stockholm, 106 91 Sweden}
\author{Yago Ferreiros}
\affiliation{Department of Physics, KTH Royal Institute of Technology, Stockholm, 106 91 Sweden}
\affiliation{IMDEA Nanociencia, Faraday 9, 28049 Madrid, Spain}
\author{Alberto Cortijo}
\affiliation{Instituto de Ciencia de Materiales de Madrid, CSIC, Cantoblanco, 28049 Madrid, Spain}
\author{Jens H.\ Bardarson}
\affiliation{Department of Physics, KTH Royal Institute of Technology, Stockholm, 106 91 Sweden}

\begin{abstract}

A space-time dependent node separation in Weyl semimetals acts as an axial vector field. 	
Coupled with domain wall motion in magnetic Weyl semimetals, this induces axial electric and magnetic fields localized at the domain wall.
We show how these fields can activate the axial (chiral) anomaly and provide a direct experimental signature of it.
Specifically, a domain wall provides a spatially dependent Weyl node separation and an axial magnetic field $\bm{B}_5$, and domain wall movement, driven by an external magnetic field, gives the Weyl node separation a time dependence, inducing an axial electric field $\bm{E}_5$.
At magnetic fields beyond the Walker breakdown, $\bm{E}_5\cdot\bm{B}_5$ becomes nonzero and activates the axial anomaly that induces a finite axial charge density---imbalance in the number of left- and right-handed fermions---moving with the domain wall.
This axial density, in turn, produces, via the chiral magnetic effect, an oscillating current flowing along the domain wall plane, resulting in a characteristic radiation of electromagnetic waves emanating from the domain wall.
A detection of this radiation would constitute a direct measurement of the axial anomaly induced by axial electromagnetic fields. 
 
 \end{abstract}

\maketitle
\textit{Introduction.}---The smallest number of Weyl fermions realizable as quasiparticles in a crystal is two~\cite{Nielsen:1981ea,Nielsen:1981ke}---one left-handed and one right-handed.
In the presence of inversion symmetry, we can choose the origin of momentum space such that one Weyl fermion resides at $\bm{b}$ and the other at $-\bm{b}$.
Since time reversal does not change the handedness of a Weyl fermion, such a minimal Weyl semimetal necessarily breaks time-reversal symmetry~\cite{Wan:2011hi,Armitage:2018dg}.
The Weyl node splitting $2\bm{b}$ is then induced by the time-reversal breaking and can be thought of as a magnetization.
Such a magnetic Weyl semimetal was recently realized in EuCd$_2$As$_2$ at intermediate temperatures~\cite{Maeaaw4718,Soh:2019cp} and in EuCd$_2$Sb$_2$ in an external magnetic field~\cite{Su:2019vg}; several further Weyl states in magnetic materials were experimentally observed~\cite{Suzuki:2019kw,Belopolski:2019bu,Liu:2019fz,Morali:2019ef}.

The electronic response of the Weyl fermions to external electromagnetic fields is fundamentally influenced by the chiral anomaly~\cite{Adler:1969ir,Bell:1969cz}.
The handedness of the Weyl fermions is not generally conserved and the axial density $n_5 = n_L-n_R$, the difference in density of left- and right-handed Weyl fermions, instead satisfies the homogeneous anomaly equation~\footnote{Throughout this work, this consistent version of the anomaly is the relevant one, cf. Refs.~\cite{Landsteiner:2016kl,Behrends:2019dk}.}
\begin{equation}
\label{chiralanomalyintro}
\partial_t n_5=\frac{e^2}{2\hbar^2\pi^2}\left( \bm{E}\cdot \bm{B}+\frac{1}{3}\bm{E}_5\cdot \bm{B}_5\right).
\end{equation}
Here $\bm{E}$ and $\bm{B}$ are the usual electric and magnetic fields, while $\bm{E}_5$ and $\bm{B}_5$ are so-called axial electric and magnetic fields~\cite{VKG10,Ilan:2019uq}, which point in the opposite direction for the two chiralities.
Direct experimental signatures of the anomaly have proven hard to come by.
While negative magnetoresitance is a consequence of the chiral anomaly~\cite{Nielsen:1983ce,Son:2013jz} it is not an unambiguous signature of it~\cite{Xiong:2015kl,Goswami:2015bs,Arnold:2016hl,Andreev:2018ei,Liang:2018cj}.
Axial fields are also challenging to realize as they may require systematic and significant straining of materials~\cite{Cortijo:2015ip,Pikulin:2016bj,Grushin:2016ji}; obtaining an axial electric field $\bm{E}_5$ is particularly hard, as this requires controllable time-dependent strain.
This is because the Weyl node separation $\bm{b}$ couples to the Weyl fermions as an axial vector potential and strain gives it a space-time dependence as $\bm{b}\rightarrow \bm{b}(\bm{r},t)$.
This then gives rise to axial fields through $\bm{B}_5=\nabla\times \bm{b}$ and $\bm{E}_5=-\partial_t\bm{b}$, in analogy with how electromagnetic fields are obtained from a vector potential~\footnote{We assume inversion symmetry such that the energy difference of the Weyl nodes $b_0 = 0$.}.
\begin{figure}[t]
\includegraphics[width=\columnwidth]{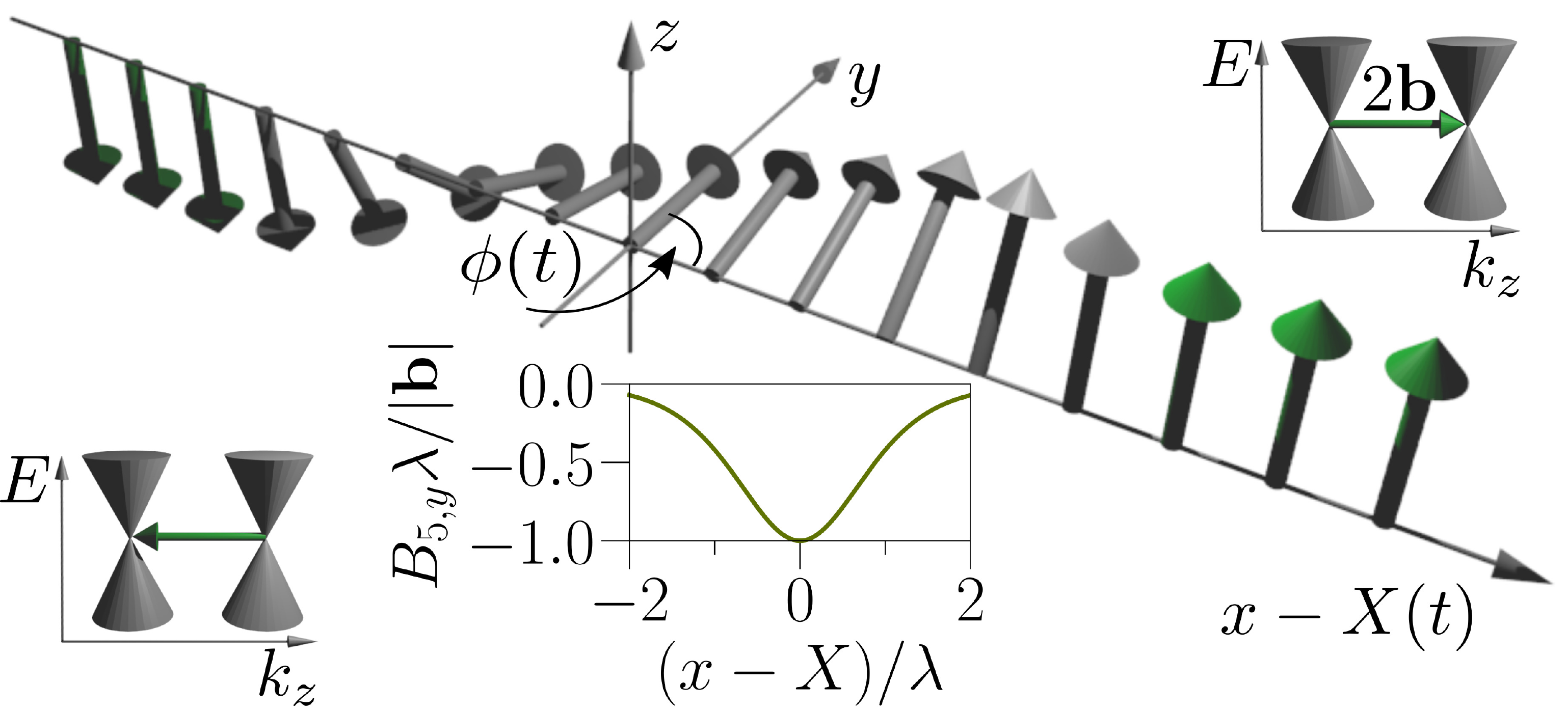}
\caption{A domain wall along the $x$-direction with a continuously varying Weyl node separation $2\bm{b}$.
The two insets with Weyl-cones show the corresponding Weyl node separation in momentum space, given by the bulk magnetization vector, (green).
The domain wall has a hard-axis anisotropy in the $y$-direction, and an easy-axis in the $z$-direction. 
$\phi(t)$ is the angle of the magnetization at the domain wall center out of the easy-axis plane ($xz$-plane), and $X(t)$ is the position of the domain wall center.
The domain wall depicted is in the Bloch configuration for which $\phi(t)=\pi/2$. 
The Bloch wall only has a nonzero component of $\bm{B}_5$ in the $y$-direction, and $B_{5,y}\lambda/|\bm{b}|$ is plotted as a function of $x-X(0)$ (middle bottom inset); for parameter values see \protect\cite{Note5}.
\label{fig:dw}
}
\end{figure}

In this work we discuss how both of these difficulties---the generation of axial fields and detection of the axial anomaly---are overcome by studying domain wall motion~\cite{Shibata:2011gz,TATARA2008213} in Weyl semimetals.
Indeed, in a magnetic Weyl semimetal a space-time variation in the Weyl node separation is naturally realized at domain walls in the magnetization~\cite{Araki:2019dz}.
Such domain walls have been indirectly observed, for example, in the magnetic nodal semimetal CeAlGe~\cite{Suzuki:2019kw}.
Domain wall motion has also been studied in related systems such as junctions of ferromagnets and topological insulators~\cite{PhysRevB.92.085416,PhysRevB.89.024413,PhysRevB.82.161401,PhysRevLett.108.187201,PhysRevB.90.041412}. 

Consider a magnetic domain wall along the $x$-direction, pointing in the $\pm z$-direction deep in the bulk, as depicted in Fig.~\ref{fig:dw}.
For concreteness, we assume the easy-axis of the magnetic anisotropy to be in the $z$-direction and the hard-axis anisotropy to lie in the $y$-direction, making the $xz$-plane the easy-plane.
The domain wall can be described in terms of two collective coordinates, the position $X(t)$ of the center of the wall and the internal angle $\phi(t)$, which measures the angle of the magnetization out of the easy-plane.
$X(t)$ and $\phi(t)$ describe zero modes of fluctuations around the domain wall arising from translation invariance along the $x$-direction and rotation invariance around the $z$-axis, respectively~\citep{rajaraman1982solitons}.
While the existence of a hard-axis anisotropy formally breaks the rotational invariance, $X(t)$ and $\phi(t)$ are still good collective coordinates in the limit of weak anisotropy.
There are two special configurations of the domain wall, the N\'{e}el wall for which $\phi=0$ where the domain wall is situated in the easy plane, and the Bloch wall, for which $\phi=\pi/2$, illustrated in Fig.~\ref{fig:dw}.
Since $\bm{b}$ rotates from $-b\hat{\bm{z}}$ to $b\hat{\bm{z}}$ an axial magnetic field localized at the domain wall is obtained.
This is similar to the $\bm{B}_5$ obtained at the surface of Weyl semimetals~\cite{PhysRevB.89.081407}, except that it is not constrained to a definite location in space.

When the domain wall moves, the magnetization becomes time-dependent, generating an axial electric field $\bm{E}_5$.
A controllable way of moving a domain wall is by a magnetic field $\bm{B}= B\hat{\bm{z}}$.
This results in a rigid shift of the domain wall center $X(t)$ with an average velocity that increases linearly with $B$ up until a critical value $B_c$, at which the internal angle starts rotating and the velocity decreases---this is called the Walker breakdown~\cite{doi:10.1063/1.1663252}.
The axial electric field generated in this movement is a function of both the rotation and the velocity of the domain wall.
However, as we show, the axial anomaly (which depends on $\bm{E}_5\cdot\bm{B}_5$) is only activated when the internal angle starts rotating, for magnetic fields larger than $B_{\rm{c}}$. 
Once it is activated an axial density $n_5$, localized at the domain wall, builds up and an oscillating current is induced, via the chiral magnetic effect~\cite{Kharzeev:2014fr}.
This results in electromagnetic radiation which is a direct signature of the axial anomaly induced by axial fields.

{\it Domain wall dynamics.}---We take the Weyl node separation in a domain wall to define a unit magnetization $\bm{m}$ as $\bm{b}(\bm{r},t)=\Delta/(e\nu_F)\bm{m}(\bm{r},t)$, where $e$ is the elementary charge, $\nu_F$ the Fermi velocity, and $\Delta$ an effective exchange coupling between the electrons and the magnetization.
The variation of $\bm{b}$ with $\bm{r}$ and $t$ is slow enough, compared to typical electronic time and length scales, that the interpretation of it as a Weyl node separation in momentum space still makes sense.
Expressed in the collective coordinates,
\begin{equation}
\label{unitm}
\bm{m}=\left(\frac{\cos[\phi(t)]}{\cosh(\frac{x-X(t)}{\lambda})},\frac{\sin[\phi(t)]}{\cosh(\frac{x-X(t)}{\lambda})},-q\tanh(\frac{x-X(t)}{\lambda})\right),
\end{equation}
where $\lambda$ is the domain wall width and $q=\pm 1$ is the topological charge~\citep{rajaraman1982solitons}; we consider the case $q=-1$, cf.~Fig.~\ref{fig:dw}. 
The dynamics of the domain wall is encapsulated in a ferromagnetic action $S_{\rm{FM}}=\int\diff t(L_{\rm{B}}-H_{\rm{H}}-H_{\rm{Z}})$ which considers the precession and exchange coupling of the magnetization, coupled to an external magnetic field~\cite{Shibata:2011gz}.
The Lagrangian describing the precession is given by a Berry phase term $L_{\rm{B}}=\hbar/a^3\int\diff^3 x\hspace{0.1cm}\dot{\phi}(\cos\theta-1)$, where $\theta=2\tan^{-1}\text{exp}[-(x-X(t))/\lambda]$ and $a$ is the lattice constant~\cite{BrokenSymmetriesSchakel}.
The exchange coupling contributes the term $H_{\rm{H}}= 1/(2a^3)\int\diff^3 x(Ja^2|\nabla\bm{m}|^2-Km_z^2+K_\perp m_y^2)$, which is the Heisenberg Hamiltonian in the continuous limit.
Here $J$, $K$ and $K_\perp$ are positive constants: $J$ is the exchange energy, and $K$ and $K_\perp$ are the easy- and hard-axis anisotropy energies.
The contribution from an external magnetic field $\bm{B}=B\hat{\bm{z}}$, applied in the direction of the easy-axis anisotropy, is included as a Zeeman term, $H_{\rm{Z}}=\hbar/a^3\int\diff^3 x\hspace{0.1cm}\bm{m}\cdot\gamma\bm{B}$, where $\gamma$ is the electron gyromagnetic ratio.

The collective coordinate description of the domain wall in terms of $X(t)$ and $\phi(t)$ is valid as long as there is translational invariance in the $x$-direction and rotational invariance around the $z$-direction.
While the existence of a hard-axis anisotropy would deform the domain wall and break the rotational invariance, the deformation is negligible in the limit $K_\perp\ll K$, in which $X(t)$ and $\phi(t)$ are good collective coordinates~\cite{TATARA2008213}.
While this is not an essential limit, it simplifies our discussion so we assume it henceforth.
The domain wall action in this limit in terms of collective coordinates~\cite{Shibata:2011gz}
\begin{equation}
\label{SFM_Xphi}
S_{\rm{FM}}=-\frac{2\hbar A}{a^3}\int \diff t\hspace{0.1cm}\left(\dot{\phi}X+\nu_\perp\sin^2\phi-\gamma B\hspace{0.1cm}X\right).
\end{equation}
Here $A$ is the cross-section of the sample in the $yz$-plane and $\nu_\perp=\lambda K_\perp/(2\hbar)$, with $\lambda=\sqrt{J/K}$ the domain wall width. 
The first term is the Berry phase term, the second the contribution from the Heisenberg Hamiltonian and the last the Zeeman term.
\begin{figure}[bt!]
\includegraphics[width=\columnwidth]{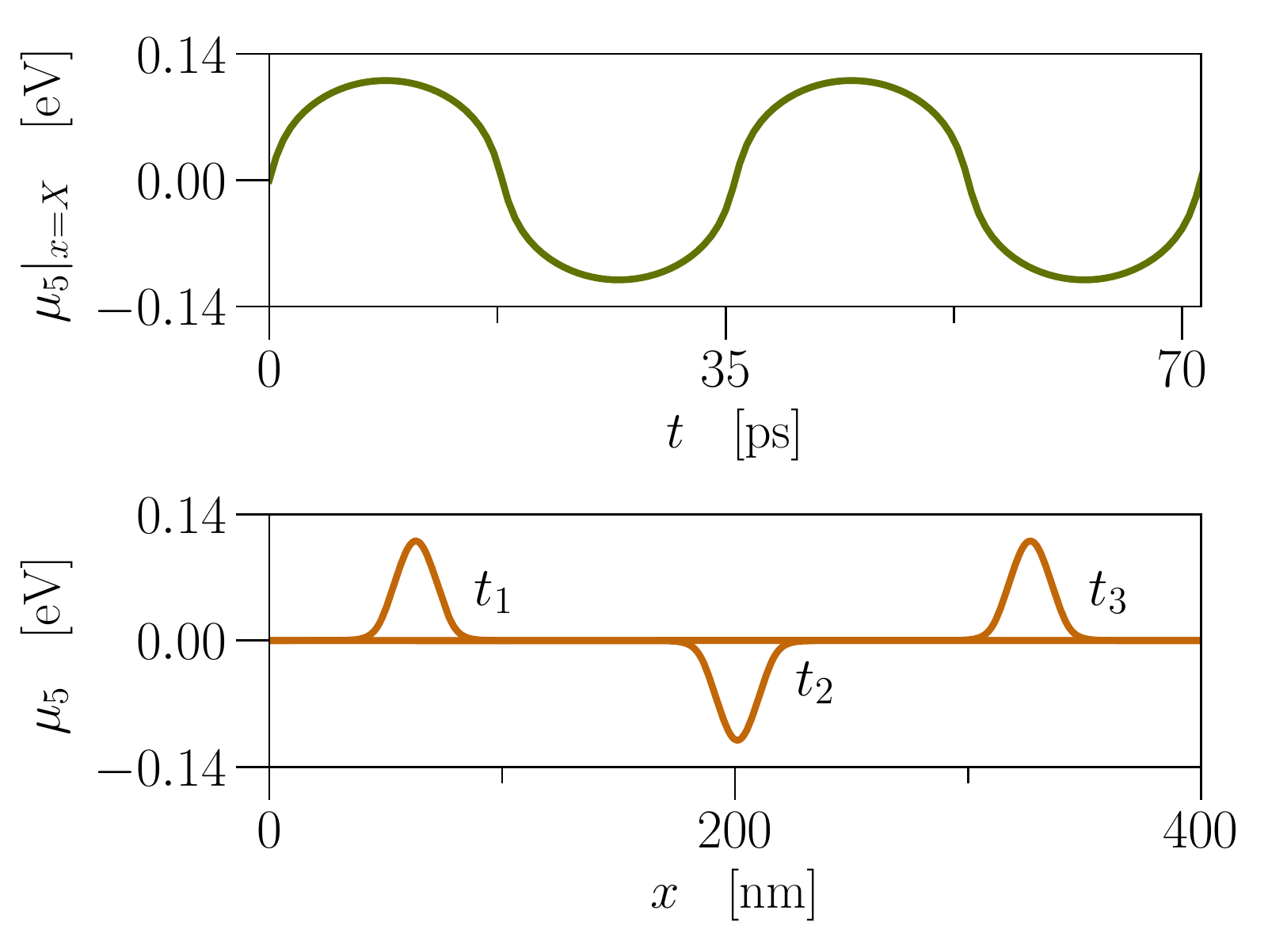}
\caption{\label{fig:mu5} Axial chemical potential $\mu_5$ as function of time $t$, evaluated at the domain wall center $x=X(t)$ (upper panel) and as a function of $x$ at times $t_1=100\tau_\phi+\tau_\phi/4$, $t_2=320\tau_\phi-\tau_\phi/4$ and $t_3=520\tau_\phi+\tau_\phi/4$ (lower panel). Here we take chemical potential $\mu=10$ meV, temperature $T=300$~K and magnetic field $B=1$ T; other parameters are given in \protect\cite{Note5}.
}
\end{figure}
The time evolution of the collective coordinates is given by the action $S_{\rm{FM}}$ together with damping, which takes into account magnetization relaxation effects.
Incorporating the damping as a dissipation function $W=-\hbar A \lambda\alpha/a^3[(\dot{X}/\lambda)^2+\dot{\phi}^2]$, where $\alpha$ is the Gilbert damping constant~\cite{Gilbert}, the generalized Euler-Lagrange equations of motion take the form
\begin{align}
\label{eom1}
&\dot{\phi} +\frac{\alpha}{\lambda}\dot{X}=\gamma B,\\
\label{eom2}
&\dot{X}-\alpha\lambda\dot{\phi}=\nu_\perp\sin 2\phi,
\end{align}
which are combined into a single equation for the internal angle: $\dot{\phi}=a_1-a_2\sin(2\phi)$. 
Here $a_1=\gamma B /\left(\alpha ^2+1\right)$ and $a_2=\alpha \nu_\perp/[(\alpha ^2+1)\lambda]$.
$a_2$ is always positive whilst the sign of $a_1$ depends on the direction of the magnetic field. 

The solutions for $\phi(t)$ depend on the magnitude of $B$ and are divided into two domains separated by the critical magnetic field $B_{\rm{c}}=\alpha  \nu_\perp/(\gamma \lambda) $ obtained when $|a_1|=a_2$.
This is observed from the solution for $\phi(t)$, which for initial condition $\phi(0)=0$, is 
\begin{equation}
\tan(\phi)=\frac{a_1\tan(\sqrt{a_1^2-a_2^2}\hspace{0.1cm}t)}{\sqrt{a_1^2-a_2^2}+a_2\tan(\sqrt{a_1^2-a_2^2}\hspace{0.1cm}t)}.
\end{equation}
The square root in the above expression is negative when $ B<B_{\rm{c}}$, in which case the solution is $\tan(\phi)=a_1\tanh(\zeta t)/[\zeta+a_2\tanh(\zeta t)]$, where $\zeta=\sqrt{a_2^2-a_1^2}$. In the long-time limit $t\rightarrow\infty$, this results in a constant angle $\phi=1/2\arcsin(B/B_{\rm{c}})$.
The domain wall velocity is also constant in this limit: $\dot{X}= \lambda\gamma B/\alpha$.
When the magnetic field is larger than the critical value, $B>B_{\rm{c}}$, the internal angle oscillates in time according to $\phi(t)=\arctan\{a_1\tan(\omega t)/[\omega+a_2\tan(\omega t)]\}$, with the angular frequency $\omega=\sqrt{a_1^2-a_2^2}$.
In this regime the domain wall position $X(t)=(-\phi(t)+\gamma Bt)\lambda/\alpha$ also increases with time with an oscillatory motion.
The magnitude of the magnetic field therefore plays a role in how the domain wall moves, which has implications for the onset of the chiral anomaly.
The anomaly equation, Eq.~\eqref{chiralanomalyintro}, (with $\bm{E}=0$) is proportional to 
\begin{equation}
\bm{E}_5\cdot\bm{B}_5=\frac{\Delta^2}{e^2\nu_F^2\lambda}\frac{\dot{\phi}\cos\phi}{\text{cosh}^3\left(\frac{x-X(t)}{\lambda}\right)},
\end{equation}
which is zero when $B<B_{\rm{c}}$, implying that the chiral anomaly is only activated in the Walker breakdown regime $B > B_{\rm{c}}$.
The axial electric field, which contributes with the term $\dot{\phi}$, is also nonzero before the Walker breakdown, but is then orthogonal to $\bm{B}_5$.

{\it The axial chemical potential.}---The axial anomaly induced by the domain wall motion generates an axial chemical potential $\mu_5=(\mu_L-\mu_R)/2$, with $\mu_L$ and $\mu_R$ the chemical potentials of left- and right-handed Weyl fermions, respectively.
The anomaly equation is of the form $\partial_t n_5=e^2/(6\hbar^2\pi^2)\bm{E}_5\cdot \bm{B}_5-n_5/\tau$, where the second term takes into account inter-valley scattering between the two Weyl cones, with inter-valley scattering time $\tau$~\cite{PhysRevX.4.031035,Behrends:2016bc} and where $\bm{E}_5\cdot \bm{B}_5$ oscillates in time with period $\tau_\phi = 2\pi/\omega$.
In the limit $\tau_\phi\ll\tau$, the domain wall oscillates faster than the inter-valley scattering and the number density becomes $n_5=e^2/(6\hbar^2\pi^2)\int_0^t\diff s\hspace{0.1cm} \bm{E}_5(x,s)\cdot \bm{B}_5(x,s)$.

The axial chemical potential is considered to be space and time dependent and relates to the axial number density as 
\begin{equation}
\begin{aligned}
\label{mu5}
\mu_5(x,t)&=\frac{2^{\frac{2}{3}}\left(\left[C_1n_5+\sqrt{C_1^2n_5^2+4C_2^3}\right]^{\frac{2}{3}}-2^{\frac{2}{3}}C_2\right)}{6\left(C_1n_5+\sqrt{C_1^2n_5^2+4C_2^3}\right)^{\frac{1}{3}}},\\
\end{aligned}
\end{equation}
where the constants $C_1=81\pi^2\hbar^3\nu^3_F$ and $C_2=3(3\mu^2+\pi^2T^2k_B^2)$, $T$ is the temperature and $k_B$ the Boltzmann constant.
The above expression for $\mu_5$ holds in the limit of small magnetic fields, $\hbar eB\ll\mu_5^2/\nu_F^2$, where $\mu=(\mu_L+\mu_R)/2$ is the average chemical potential~\cite{PhysRevD.78.074033}.
$\mu_5(x,t)$ oscillates in time, is located at the domain wall, and travels along the $x$-direction as $X(t)$ evolves with time, see Fig.~\ref{fig:mu5}.

{\it Measuring the anomaly.}---The axial chemical potential generates a current density~\footnote{There is an additional axial contribution to the chiral magnetic effect $\bm{J}=e^2/(2\pi^2\hbar^2)\mu\bm{B}_5$~\cite{Landsteiner:2016kl}; this does not affect our results as $B_z(x,t)$ is odd around the domain wall center and thus averages to zero.} proportional to the external magnetic field through the chiral magnetic effect~\cite{PhysRevD.78.074033},
\begin{equation}
\label{cme}
\bm{J}^A(x,t)=\frac{e^2}{2\pi^2\hbar^2}\mu_5(x,t)B\hat{\bm{z}} \equiv J^A_z(x,t)\hat{\bm{z}}.
\end{equation}
The rotation of the magnetization further yields an effective current, $\bm{J}^M(x,t)=\nabla\times \bm{M}$, where $\bm{M}=\gamma\hbar/a^3\bm{m}$ is the magnetization density.
The currents $J^A_z(x,t)$ and $\bm{J}^M(x,t)$ give rise to electromagnetic fields, measurable through their radiated power.
We describe these fields using Jefimenko's equations~\cite{Jackson}, and consider them separately in the near-field $r\ll R_0$, and the far-field $r\gg R_0$, limits, where $R_0$ is the wavelength of the electromagnetic fields and $r=\sqrt{[x-X(t)]^2+y^2+z^2}$ is the distance to the detector from the domain wall center~\footnote{The distance from the domain wall to the detector is $R=|\bm{r}-\bm{r}'|$, where $\bm{r}$ is the detector coordinates and $ \bm{r}'$ is the domain wall coordinates. We assume the limit $r'\ll r$, such that $R\sim r$, and therefore simply refer to the distance to the detector as $r$.}.
In both limits the domain size is considered to be the smallest of the three length scales, $\lambda\ll r,R_0$. 
We further require that $L_z>\nu_F\tau_\phi$, where $L_z$ is the width of the sample in the $z$-direction (in the opposite limit accumulation of charge at the edge of the sample might become relevant), which for realistic parameters~\footnote{Lattice constant $a=0.5$ nm, hard axis anisotropy $K_\perp/a^3=10^2$ J/m$^3$, domain wall width $\lambda=10$ nm, domain wall length in $y$-, $z$-directions, $L_y=L_z=10\mu$m, Fermi velocity $\nu_F=5\cdot 10^5$ m$/$s, inter-valley scattering rate $\tau= 1$ ns, Gilbert damping constant $\alpha=0.01$ and half the length of the Weyl node separation $|\bm{b}|=0.1\pi/a$} and $B\gtrsim1$~T holds when $L_z\sim 10$ $\mu$m.
The radiation in the near-field is measurable by current-technology on-chip, which can detect weak signals of only a few emitted photons~\cite{Gustavsson:2008gb}.
Reactive components dominate the electromagnetic fields in the near field, and by describing the electromagnetic fields in this limit as an expansion in $r/c$, we find that the only radiative contribution up to second order in $r/c$ originates from the anomaly current. 
The resulting power impinging on a detector of size $\ell\times\ell$, and a small solid angle, is given by $P^A_\ell(\theta,\varphi)=\ell^2 d_2^2B^2\lambda^2\langle(\partial_t^2\mu_5)^2\rangle\sin^2\theta/(2\varepsilon_0c^5)$, where $d_2=L_yL_ze^2/(8\pi^3\hbar^2)$, $\varepsilon_0$ is the vacuum permeability, $\theta$, $\varphi$ are the polar and azimuthal angles and $\langle...\rangle$ refers to time average (see the Appendix for details).
Fig.~\ref{fig:Efield} depicts the number of photons, $n=P^A_{\ell}t_\ell/E_p$, emitted by a domain wall during the time $t_\ell=\ell/\langle\dot{X\rangle}$ it takes it to traverse the detector length $\ell=1$ $\mu$m, as a function of the external magnetic field, where $E_p= \hbar\omega$ is the photon energy. 
The number of emitted photons goes above 1 for $B\sim 2.5$ T and increases as $B^4$, yielding around $1530$ emitted photons for $B=15$ T, making the signal observable by detector sizes $\ell\sim0.1-1$ $\mu$m~\cite{Aguado:2000ih,Gustavsson:2006jm,Gustavsson:2008gb,Govenius:2016fo}.

In the far-field limit the electromagnetic fields radiate, and the power contains contributions from both $J^A_z(x,t)$ and $\bm{J}^M(x,t)$.
The anomaly current is even across the domain wall, whilst the magnetzation current is odd, which results in the radiation due to the magnetization decaying faster with distance than the radiation due to the anomaly (total power due to the anomaly is independent of $r$ while the magnetization power decays as $r^{-2}$).
For $B\sim 10$ T and $r\sim 1$ cm the contribution to the power due to the anomaly is $10^4$ times larger than the magnetization contribution (for details see the Appendix).
The power in both limits is thus dominated by the anomaly contribution, and the radiation frequencies range between $\omega/2\pi=27-420$ GHz for magnetic fields in the interval $B=1-15$ T.

\begin{figure}[bt!]
\includegraphics[width=\columnwidth]{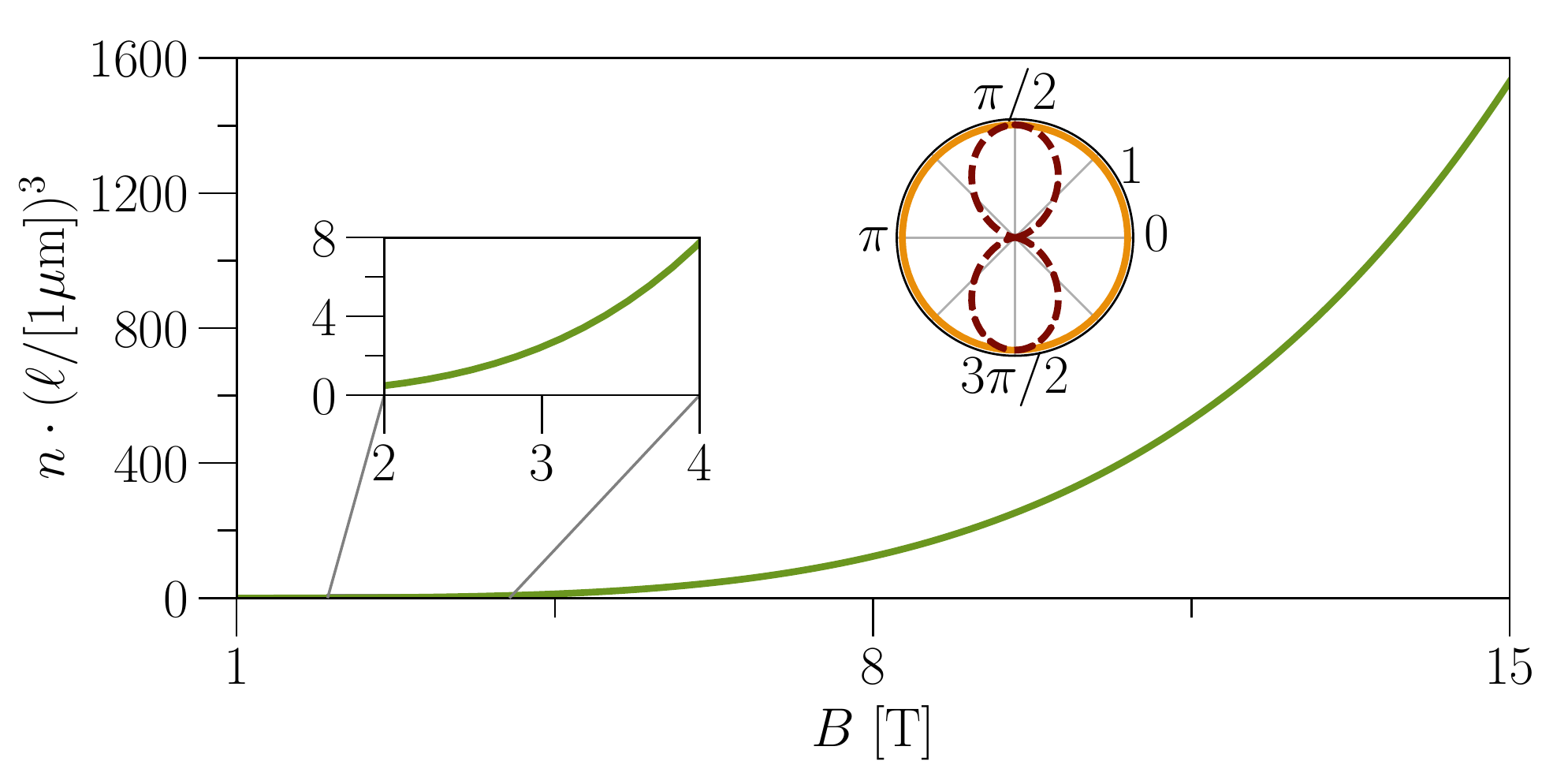}
\caption{\label{fig:Efield} The number of photons $n=P^A_{\ell,\rm{max}}t_\ell/E_p$ emitted by the domain wall during the time $t_\ell=\ell/\langle\dot{X\rangle}$ it traverses the detector length $\ell=1\mu$m, as a function of external magnetic field $B$. Here $P^A_{\ell,\rm{max}}=\ell^2 d_2^2B^2\lambda^2\langle(\partial_t^2\mu_5)^2\rangle\sin^2\theta/(2\varepsilon_0c^5)|_{\theta=\pi/2}$ is the radiated power in the near-field, and is due to the anomaly current, $E_p=\hbar\omega$ is the photon energy.
The polar plot depicts the normalized power per solid angle, $(\diff P^A/\diff\Omega)/\mathcal{A}=\sin^2\theta$, where $\mathcal{A}$ is the amplitude of $(\diff P^A/\diff\Omega)$, as a function of the polar angle $\theta$ (maroon, dotted) for any value of the azimuthal angle $\varphi$, and as a function of $\varphi$ (orange, solid) for $\theta=\pi/2$.
Here $L_y=L_z=10$  $\mu$m, for other parameter values see Fig.~\ref{fig:mu5} and~\protect\cite{Note5}.
}
\end{figure}

These results generalize to multiple domain walls.
Adjacent domain walls have opposite topological charge $q=\pm 1$, and therefore travel in opposite directions under the influence of a magnetic field.
The current due to the anomaly is independent of the topological charge as the axial chemical potential goes as $q^2$.
This implies that the radiated field from the anomaly current of two adjacent domain walls add up.
The magnetization current on the other hand has a $q$-dependence which produces a sign different in the $y$-components of the magnetization current at two adjacent domain walls, which modifies the radiation when considering several domain walls compared to that of a single domain wall.
In any case, this will not change the fact that the anomaly contribution is the dominating one, and measurable also in the case of multiple domain walls.
However, the radiated power only exists during a finite length of time depending on the velocity of the domain walls, until the domain walls annihilate each other or until they reach the boundary of the sample.
Pinning~\cite{TATARA2008213}---local enhancement of easy-axis anisotropy, due to for example impurities, confining the domain wall to a certain region---could modify the details of the radiation field, since adjacent domain walls could be prevented from annihilating one another and the electromagnetic radiation would come from a fixed location.

{\it Discussion.}---We have shown how field-driven motion of a domain wall in a magnetic Weyl semimetal leads to the activation of the axial anomaly.
This results from the space and time dependent Weyl node separation emerging from the domain wall motion, which generates axial electromagnetic fields.
The anomaly generates an axial chemical potential at the domain wall, which in turn results in an oscillating current and electromagnetic microwave radiation, detection of which would constitute a direct measurement of the axial anomaly.
Experimental techniques to detect such microwave radiation are advanced and can even be done on-chip~\cite{Aguado:2000ih,Gustavsson:2006jm,Gustavsson:2008gb,Govenius:2016fo}. 
While we have made some simplifying approximations to highlight the fundamental physics, we expect the qualitative picture to be robust in realistic situations, and a general feature of any domain wall motion in Weyl semimetals.
For example, current-driven domain wall motion will lead to the same axial anomaly-triggering mechanism as the one described here, but will allow for an electronic control of anomaly activation, which may be useful in designing experiments and applications. 
We have also worked in the limit of weak hard-axis anisotropy where a description of the domain wall in terms of collective coordinates is sufficient. 
Deviations away from this limit will lead to a more complicated theory that needs to take into account modes beyond just the zero modes we include, but this is not expected to modify the qualitative description of the emergence of axial fields located at the domain wall.

We have focused our discussion on the use of domain wall motion for detecting anomaly physics.
The other way around, namely the effects of the anomaly on the physics of domain walls and related spintronics phenomena is an interesting avenue for future studies.\\

\acknowledgements
This work was supported by the ERC Starting Grant No.~679722. Y.F.\ acknowledges financial support through the Programa de Atracción de Talento de la Comunidad de Madrid, Grant No. 2018-T2/IND-11088. A.C.\ acknowledges financial support through MINECO/AEI/FEDER, UE Grant No. FIS2015-73454-JIN and European Union structural funds and the Comunidad Autonoma de Madrid (CAM) NMAT2D-CM Program (S2018-NMT-4511).

\appendix
\section{Appendix A: Radiation in the near and far field}
\renewcommand{\theequation}{A.\arabic{equation}}
\setcounter{equation}{0}

The presence of the chiral anomaly induces a current through the chiral magnetic effect, acting  as a source of radiating electromagnetic fields.
The curl of the magnetization yields an effective current which contributes to the radiating  electromagnetic fields.
Here we discuss the radiated fields due to the two currents in the near-field and far-field limits, and show that the radiated power is dominated by the contribution due to the chiral anomaly.

The source of electromagnetic radiation enters Maxwell's equations~\cite{Jackson} as the right hand side of
\begin{align}
\label{eq:maxwell}
\nabla\times\frac{1}{\mu_0}\bm{B}-\varepsilon_0\frac{\partial \bm{E}}{\partial t}&=\bm{J}^A+\nabla\times \bm{M},
\end{align}
where $\varepsilon_0$ and $\mu_0$ are the vacuum permittivity and vacuum permeability respectively, and
\begin{align}
\bm{J}^{A}&=\frac{e^2}{2\pi^2\hbar^2}\mu_5(x,t)\bm{B}_{\rm{ext}}
\end{align}
is the current generated by the chiral anomaly through the chiral magnetic effect where $\mu_5(x,t)$ is the axial chemical potential and $\bm{B}_{\rm{ext}}$ is the applied magnetic field.
The second source term in Eq.~(\ref{eq:maxwell}) is an effective current generated through the curl of the magnetization per unit volume, $\bm{M}=\gamma\hbar/a^3\bm{m}$.
We refer to this effective current contribution as a magnetization current and define it as
\begin{align}
\bm{J}^{M}&=\nabla\times \bm{M}.
\end{align}
The total source of electromagnetic fields contained at the domain wall is therefore $\bm{J}=\bm{J}_{M}+\bm{J}_{A}$. 

The currents give rise to electromagnetic fields, described by Jefimenko's equations as~\cite{Jackson}:
\begin{equation}
\label{E}
\bm{E}(\bm{r},t)=-\frac{1}{4\pi\varepsilon_0 c^2}\int\diff r'\frac{\left[\partial_t\bm{J}(\bm{r}',t')\right]_{\rm{ret}}}{R},
\end{equation}
\begin{equation}
\begin{aligned}
\label{B}
\bm{B}(\bm{r},t)&=\frac{\mu_0}{4\pi}\int\diff r'\left( \frac{\left[ \bm{J}(\bm{r}',t')\right]_{\rm{ret}}\times \hat{\bm{R}}}{R^2}\right.\\
&\left.\hspace{2cm}+ \frac{\left[\partial_t \bm{J}(\bm{r}',t')\right]_{\rm{ret}}\times\hat{\bm{R}}}{cR}\right),
\end{aligned}
\end{equation}
where $\bm{R}=\bm{r}-\bm{r}'$, such that $R=|\bm{R}|$ is the distance from the source, and $[...]_{\rm{ret}}$ means that the quantity is to be evaluated at the retarded time $t_r=t-r/c$.
The fields are evaluated by considering the three relevant length scales, $\lambda$ the domain wall width, $R_0$ the wavelength of the radiation, and $R$.
The limit $R\gg R_0$ constitutes the far-field and $R\ll R_0$ the near field; in both cases we assume the limit $r'\sim \lambda\ll r$, where the extent of the region containing the currents, $r'$, is much smaller than $r$ and is always the smallest length scale.
The wavelength $R_0=\tau_\phi c$ decreases as $B^{-1}$ through the period of the current $\tau_\phi=2\pi/\omega$.
For $B=1$ T, $R_0\sim 10$ mm, while for $B=15 $ T $R_0\sim 0.7$ mm.
The size of the domain in the $y$- and $z$- directions, $L_y$ and $L_z$, is also considered to be small in comparison to $r$, where the width of $L_z>\nu_F\tau_\phi$ is large enough so that the oscillating current does not start accumulating charge at the edge.
The behavior of the radiated fields in the far- and near-field limits differ in character: in the far-field the fields are radiative, and in this region the total radiated power is well defined.
In the near-field limit Jefimenko's equations are expanded in terms of $R/R_0$.
In this limit the fields are dominated by reactive fields, where the radiative components that appear are higher order in $R/R_0$ than the former and thus less prominent~\cite{Gregson:2007}.
Even so, we first proceed by exploring the near-field to show the qualitative appearance of a radiated power due to the chiral anomaly.

\textit{Radiation in the near-field.}---In the instantaneous approximation, or near-field, where  $R\ll R_0$, we have the relation $|t-t_r|/\tau_\phi=R/R_0\ll 1$, such that the radiated fields in Eqs.~(\ref{E})--(\ref{B}) given in retarded time may be expanded in terms of $|t-t_r|=R/c$, which to the lowest orders yields
\begin{align}
\label{Enf}
\bm{E}(r,t)&=-\frac{1}{4\pi\varepsilon_0}\int\left(\frac{\partial_t\bm{J}}{c^2R}-\frac{\partial_t^2\bm{J}}{c^3}\right)\diff\bm{r}',\\
\label{Bnf}
\bm{B}(r,t)&=\frac{\mu_0}{4\pi}\int\left(\frac{\bm{J}\times \bm{R}}{R^3}-\frac{(\partial_t^2\bm{J})\times \bm{R}}{2c^2R}\right)\diff\bm{r}',
\end{align}
for the electromagnetic fields.

As we are considering the limit $\lambda\ll r$ we may further expand in $r^{-1}$, keeping only the lowest order terms, for example
\begin{equation}
\frac{1}{R}=\frac{1}{r}\left(1+\frac{2\left[x_q\tilde{x}+yy'+zz'\right]}{r^2}+\mathcal{O}\left[\frac{1}{r^2}\right]\right),
\end{equation}
where $\tilde{x}=x'-X(t)$ and $x_q=x-X(t)$, such that $r=\sqrt{x_q^2+y^2+z^y}$ is defined as the distance from the domain wall center.
The anomaly contribution to the current is proportional to $\mu_5(x,t)$, and its time derivatives,  where the axial chemical potential is given by
\begin{equation}
\begin{aligned}
\label{mu5}
\mu_5(x,t)&=\frac{2^{\frac{2}{3}}\left(\left[C_1n_5+\sqrt{C_1^2n_5^2+4C_2^3}\right]^{\frac{2}{3}}-2^{\frac{2}{3}}C_2\right)}{6\left(C_1n_5+\sqrt{C_1^2n_5^2+4C_2^3}\right)^{\frac{1}{3}}},\\
\end{aligned}
\end{equation}
where the constants $C_1=81\pi^2\hbar^3\nu^3_F$ and $C_2=3(3\mu^2+\pi^2T^2k_B^2)$.
The axial density in the limit $\tau_\phi \ll \tau$ is 
\begin{equation}
n_5(x,t)=\int_0^t \bm{E}_5\cdot\bm{B}_5 \diff s =\frac{\Delta^2}{e^2\nu_F^2\lambda}\int_0^t \frac{\dot{\phi}\cos\phi}{\cosh^3\left(\frac{x-X(s)}{\lambda}\right)} \diff s.
\end{equation}
To simplify the evaluation of the electromagnetic fields we note that since $\bm{E}_5\cdot\bm{B}_5$ peaks at the center of the domain wall, the value of the axial density may be approximated by its value at $x=X(t)$.
This requires that $\tau_\phi\ll \lambda/\langle\dot{X}\rangle$, namely the timescale of building up a nonzero axial density must be much smaller than the time the it takes the domain wall to move away.
In this limit the axial density becomes
\begin{equation}
n_5(X(t),t) = \frac{\Delta^2}{e^2\nu_F^2\lambda}\int_0^t\frac{\dot{\phi}\cos\phi}{\cosh^3\left(\frac{X(t)-X(s)}{\lambda}\right)} \diff s,
\end{equation}
which, as the nonzero contributions from the integrand come from $s\sim t$, is approximately equal to
\begin{equation}
n_5(X(t),t) = \frac{\Delta^2}{e^2\nu_F^2\lambda}\sin\phi.
\end{equation}

Assuming an external magnetic field in the $z$-direction, $\bm{B}_{\rm{ext}}=B\hat{\bm{z}}$, together with the above consideration, yields an electric field in the $z$-direction:
\begin{align}
E_z^A(\bm{r},t)&=-\frac{d_2}{\varepsilon_0c^2}B\lambda\left(\partial_t\mu_5(X(t),t)\frac{1}{r}-\frac{\partial_t^2\mu_5(X(t),t)}{c}\right),
\end{align}
as well as components of magnetic fields in the $x$-, and $y$-directions:
\begin{align}
B_x^A(\bm{r},t)&=-d_2\mu_0 B\lambda \left(\mu_5(X(t),t)\frac{y}{r^3}-\frac{\partial_t^2\mu_5(X(t),t)}{2c^2}\frac{y}{r}\right),\\
B_y^A(\bm{r},t)&=d_2\mu_0 B\lambda \left(\mu_5(X(t),t)\frac{x_q}{r^3}-\frac{\partial_t^2\mu_5(X(t),t)}{2c^2}\frac{x_q}{r}\right).
\end{align}
Here the superscript $A$ refers to the source coming from the anomaly and $d_2=L_yL_ze^2/(8\pi^3\hbar^2)$.
Similarly, the electromagnetic fields due to the magnetization current, are
\begin{align}
E_y^M(\bm{r},t)&=-\frac{d_1}{\varepsilon_0c^2}2q\dot{X}\frac{x_q}{r^3},\\
E_z^M(\bm{r},t)&=\frac{d_1}{\varepsilon_0c^2}\pi\dot{\phi}\cos\phi\lambda\frac{x_q}{r^3},
\end{align}
and
\begin{equation}
\begin{aligned}
B_x^M(\bm{r},t)&=d_1\mu_0\left(2q\frac{z}{r^3}+\pi\sin\phi\lambda\frac{3x_qy}{r^5}-q \ddot{X} \frac{x_qz}{c^2r^3}\right.\\
&+\left.\pi\lambda\left[\dot{\phi}^2 \sin\phi -\ddot{\phi} \cos\phi\right]\frac{x_qy}{2c^2r^3}\right),
\end{aligned}
\end{equation}
\begin{equation}
\begin{aligned}
B_y^M (\bm{r},t)&=d_1\mu_0\left(-\pi\sin\phi\lambda\frac{\left[3x_q^2-r^2\right]}{r^5}\right.\\
&\left.-\pi\lambda\left(\dot{\phi}^2 \sin\phi -\ddot{\phi} \cos\phi\right)\frac{\left[x_q^2-r^2\right]}{2c^2r^3}\right),
\end{aligned}
\end{equation}
\begin{align}
B_z^ M(\bm{r},t)&=-d_1\mu_0\left(2q\frac{x_q}{r^3}-2q \ddot{X}\left[x_q^2-r^2\right]\frac{1}{2c^2r^3}\right).
\end{align}
The superscript $M$ assigned to the fields refers to the magnetization origin of the current, and $d_1=L_yL_z\gamma\hbar/(4\pi a^3)$.

Electromagnetic fields are dominantly reactive in the near-field, radiative contributions do appear, but these terms are higher order in the expansion in Eqs.~(\ref{Enf})-(\ref{Bnf}), which means that their amplitudes are small in comparison to the reactive fields.
The only fields that are radiative, to the order we consider, and thus contribute to a total power, are contributions coming from the anomaly current.
The time averages $\bra\mu_5\partial_t\mu_5\ket=\bra\partial_t\mu_5\partial_t^2\mu_5\ket=0$, so the time averaged Poynting vector $\bra\bm{S}^A\ket$ is composed of components:
\begin{align}
\langle S_x^A\rangle&=\frac{d_2^2}{\varepsilon_0c^5}B^2\lambda^2\langle(\partial_t^2\mu_5)^2\rangle\frac{x_q}{2r},\\
\langle S_y^A\rangle&=\frac{d_2^2}{\varepsilon_0c^5}B^2\lambda^2\langle(\partial_t^2\mu_5)^2\rangle\frac{y}{2r}.
\end{align}
The total power per solid angle is given by~\cite{Jackson}
\begin{equation}
\begin{aligned}
\frac{\diff P^A}{\diff \Omega}&=r^2\bm{n}\cdot\bra \bm{S}\ket\\
&=\frac{d_2^2}{2\varepsilon_0c^5}B^2\lambda^2\langle(\partial_t^2\mu_5)^2\rangle r^2\sin^2\theta,
\end{aligned}
\end{equation}
where $\bm{n}$ is a unit vector in the direction of $r$ and $\theta$ is the polar angle, so integrating over a surface of a sphere of radius $r$ yields the total power:
\begin{align}
P^A&=\frac{d_2^2}{\varepsilon_0c^5}\frac{4\pi}{3}B^2\lambda^2\langle(\partial_t^2\mu_5)^2\rangle r^2.
\end{align}

The power  radiated towards a detector with area $\ell^2$, assuming that the solid angle of the detector is very small, is given by
\begin{equation}
P_\ell^A=\ell^2\bm{n}\cdot\bra \bm{S}\ket,
\end{equation}
which is maximal for $\theta=\pi/2$.
The rate of photons $\dot{N}$ emitted per unit time is thus
\begin{equation}
\dot{N}=\frac{P_\ell^A}{E_p},
\end{equation}
where the photon energy $E_p=\hbar\omega$, $\omega$ being the angular frequency of the radiated fields.
The number of photons emitted by a domain wall passing by a detector of length $\ell$ is then given by
\begin{equation}
n=\dot{N}t_\ell,
\end{equation}
where $t_\ell=\ell/\langle \dot{X}\rangle$ is the time it takes the domain wall to pass the length of the detector.
The number of emitted photons depends on $B$ both through the the power $P^A_{\ell}$, which has a quartic $B$-dependence, and through $\langle \dot{X}\rangle$ which is linear in $B$.
For $\theta=\pi/2$ and $B=10$ T the number of emitted photons is of order $20$ for a detector of length $\ell=0.4$ $\mu$m and of order $300$ photons for $\ell=1$ $\mu$m.

\textit{Radiation in the far-field.}---The electromagnetic fields in the far-field approximation are \cite{Fitzpatrick:2008}:
\begin{align}
\label{EAn}
\bm{E}(\bm{r},t)&=-\frac{1}{4\pi\varepsilon_0 c^2}\frac{\left[\int\partial_t\bm{J}_\perp\diff\bm{r}'\right]}{r},\\
\label{BAn}
\bm{B}(\bm{r},t)&=\frac{\mu_0}{4\pi}\frac{\left[\int\partial_t\bm{J}_\perp\diff\bm{r}'\right]\times \bm{r}}{cr^2},
\end{align}
where  $\bm{J}_\perp=\bm{J}-(\bm{J}\cdot\bm{R})\bm{R}/R^2$.
The expressions are evaluated with the same approximations as for the near-field, namely assuming that the axial chemical potential is only non-zero at the center of the domain wall, as well as expanding all expression in terms of $r^{-1}$ to the lowest order, and where we assume the external magnetic field to be in the $z$-direction.
The radiated fields due to the anomaly takes the form
\begin{align}
E_x^A(\bm{r},t_r)&=\frac{d_2}{\varepsilon_0 c^2}\lambda B\partial_t\mu_5(X(t),t) \frac{zx_q}{r^3},\\
E_y^A(\bm{r},t_r)&=\frac{d_2}{\varepsilon_0 c^2}\lambda B\partial_t\mu_5(X(t),t)\frac{zy}{r^3},\\
E_z^A(\bm{r},t_r)&=-\frac{d_2}{\varepsilon_0 c^2}\lambda B\partial_t\mu_5(X(t),t)\left(1-\frac{z^2}{r^2}\right)\frac{1}{r},
\end{align}
for the three components of the electric field, and
\begin{align}
B_x^A(\bm{r},t_r)&=-d_2\mu_0\lambda B\partial_t\mu_5(X(t),t)\left(\frac{ z^2y}{cr^3}+\frac{y}{cr^2}\left[1-\frac{ z^2}{r^2}\right]\right),
\end{align}
\begin{align}
B_y^A(\bm{r},t_r)&=d_2\mu_0\lambda B\partial_t\mu_5(X(t),t)\left(\frac{ z^2x_q}{cr^3}+\frac{x_q}{cr^2}\left[1-\frac{z^2}{r^2}\right]\right),
\end{align}
\begin{align}
B_z^A(\bm{r},t_r)&=0,
\end{align}
for the magnetic field components.
The corresponding fields due to the curl of the magnetization are
\begin{align}
E_x^M(\bm{r},t_r)&=\frac{d_1}{\varepsilon_0 c^2}\left[2x_q^2-r^2\right]\left(q2\dot{X}\frac{y}{r^5}-\pi\cos\phi\dot{\phi}\frac{z\lambda}{r^5}\right),\\
E_y^M(\bm{r},t_r)&=\frac{d_1}{\varepsilon_0 c^2}\left(q4\dot{X}\frac{x_qy^2}{r^5}-2\pi\cos\phi\dot{\phi}\frac{x_q\lambda y z}{r^5}\right),\\
E_z^M(\bm{r},t_r)&=\frac{d_1}{\varepsilon_0 c^2}\left(q4\dot{X}\frac{x_qy z}{r^5}-2\pi\cos\phi\dot{\phi}\frac{x_qz^2\lambda}{r^5}\right),
\end{align}
and
\begin{align}
B_x^M(\bm{r},t_r)&=0,\\
B_y^M(\bm{r},t_r)&=d_1\mu_0\left(-q2\dot{X}\frac{yz}{cr^4}+\pi\cos\phi\dot{\phi}\frac{z^2\lambda}{cr^4}\right),\\
B_z^M(\bm{r},t_r)&=d_1\mu_0\left(q2\dot{X}\frac{y^2}{cr^4}-\pi\cos\phi\dot{\phi}\frac{yz\lambda}{cr^4}\right).
\end{align}

\begin{figure}[bt!]
\includegraphics[width=\columnwidth]{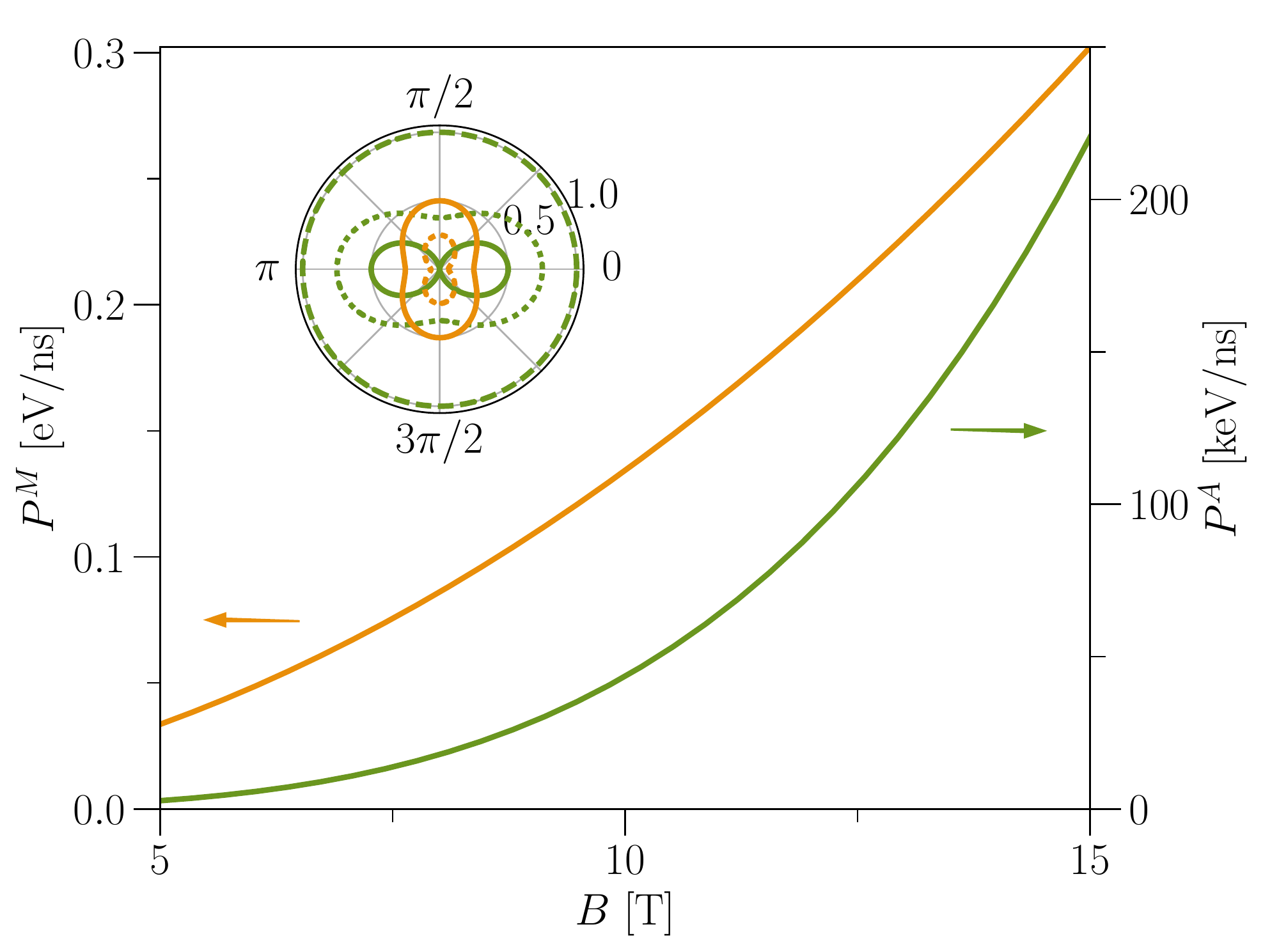}
\caption{\label{fig:P_Polar} The main plot shows the contribution from the anomaly, $P^A$, and the contribution from the magnetization, $P^M$, to the total power as functions of the applied magnetic field $B$ in the far-field limit. Note the different scale.
The inset depicts the normalized power per solid angle, $(\diff P^A/\diff \Omega)/\mathcal{A}_A$ for the anomaly contribution and $(\diff P^M/\diff \Omega)/\mathcal{A}_A$ for the magnetization contribution, where $\mathcal{A}_A$ and $\mathcal{A}_M$ are the amplitudes of the respective quantities, as functions of the azimuthal angle $\varphi$ for constant values of $\theta$, where the radius of the plot is the magnitude of the power for a given angle. Here the $--$ dashed curve corresponds to $\theta=\pi/2$, the $..$ dotted curve $\theta=\pi/3$ and the solid line, $\theta=\pi/4$. In both plots the anomaly contribution is green, and the magnetization contribution is orange. Here $r=1$ cm, for other parameter values see~\protect\cite{Note5}.
}
\end{figure}

The power per solid angle of these radiation fields takes the form:
\begin{equation}
\begin{aligned}
\frac{\diff P}{\diff \Omega}=&\frac{d_2^2}{\varepsilon_0 c^3}B^2\lambda^2\langle(\partial_t\mu_5)^2\rangle\sin^2\theta\left(1-2\sin^2\varphi\cos^2\theta\right)\\
+&\frac{d_1^2}{\varepsilon_0 c^3}\frac{1}{r^2}\left(4\langle\dot{X}^2\rangle\sin^2\varphi\sin^2\theta\left[\cos^2\theta+\sin^2\varphi\sin^2\theta\right]\right.\\
&\hspace{0.4cm}\left.+\lambda^2\pi^2\langle\cos^2\phi\dot{\phi}^2\rangle\cos^2\theta\left[\cos^2\theta+\sin^2\varphi\sin^2\theta\right]\right)\\
+&\frac{d_1d_2}{\varepsilon_0 c^3}B\langle\partial_t\mu_5\cos\phi\dot{\phi}\rangle\frac{\pi\lambda^2}{r}\cos\varphi\sin\theta\cos^2\theta\left(2\right.\\
&\left.\hspace{3.5cm}+\cos^2\theta+\sin^2\varphi\sin^2\theta\right).\\
\end{aligned}
\end{equation}
The angular dependence of the power, as a function of the azimuthal angle $\varphi$, is depicted in Fig.~\ref{fig:P_Polar} for the normalized contributions from the anomaly and the magnetization for different values of the polar angle $\theta$.
Since the magnetization term proportional to $\langle\cos^2\phi\dot{\phi}^2\rangle$ is several order larger than the term proportional to $\langle\dot{X}^2\rangle$, we only consider the radiation from the former.
The anomaly contribution is maximal for $\theta=\pi/2$ for which it has a constant $\varphi$-dependence, and it is zero at $\theta=0,\pi$.
The $\varphi$-dependence for other values of $\theta$ is such that the power at these angles is maximum for $\varphi=0,\pi$ and minimum for $\varphi=\pi/2$,
The magnetization contribution on the other hand is constant in $\varphi$ for $\theta=0,\pi$, and in between  these values of $\theta$ it is maximum for $\varphi=\pi/2$ and minimum for $\varphi=0,\pi$.
The radiation from the two contributions thus have a different angular dependence.

The terms which mix contributions from both the anomaly and the magnetization do not contribute to the total power, which is
\begin{equation}
\begin{aligned}
P(r)&=\frac{d_2^2}{\varepsilon_0 c^3}\lambda^2B^2\langle(\partial_t\mu_5)^2\rangle\frac{3\pi^2}{4}\\
&+\frac{d_1^2}{\varepsilon_0 c^3}\frac{1}{r^2}\left(\frac{13\pi^2}{8}\langle\dot{X}^2\rangle+\frac{\pi^3}{8}(\pi+3)\lambda^2\langle\cos^2\phi\dot{\phi}^2\rangle\right).
\end{aligned}
\end{equation}
Note that the contribution from the anomaly is constant in $r$, while the magnetization contribution falls off quadratically with distance. 
This stems from the fact that only terms odd in $\tilde{x}$ in $\partial_t\bm{J}^M$ contribute to the electromagnetic fields, which yields a higher order dependence in $r^{-1}$ than the contribution stemming from $\partial_t\bm{J}^A$, which is even across the domain wall.
Note also that the rotation of the internal angle, $\dot{\phi}$, and the velocity of the domain wall both are linear in the external magnetic field, and that $\partial_t\mu_5$ in turn is linear in $\dot{\phi}$.
This means that the anomaly contribution is quartic in the external magnetic field, while the magnetization contribution is merely quadratic, which renders the anomaly contribution the dominating one for larger values of $B$.
The total power as a function of the external magnetic field is displayed in Fig.~\ref{fig:P_Polar} for parameter values~\cite{Note5} and $r=1$ cm, which shows the different $B$-dependencies of the two contributions and also depicts how the anomaly contribution is several orders of magnitude larger than the contribution from the magnetization. 
For $B=5-10$ T, $R_0\sim 1$ mm, so the distance to the detector in the far-field is of order $r=1$ cm and larger which, since the power due to the anomaly is constant in $r$ whilst the magnetization contribution decays as $r^{-2}$, implies that the anomaly contribution always dominates in the far-field.
\bibliography{AxialDwLib}

\end{document}